\documentclass[aps, pre, preprint, superscriptaddress]{revtex4-1}
\usepackage{lipsum}
\usepackage{epsfig}
\usepackage{amssymb,amsmath,amsthm}
\usepackage{color}
\usepackage{textcomp}
\usepackage[colorlinks=true,linkcolor=blue,citecolor=blue]{hyperref}

\begin{document}


\title{Stability of bubble-like fluxons in disk-shaped Josephson junctions in the presence of a coaxial dipole current}



\author{Alicia G. Castro-Montes}
\email[]{alici.gabriela@gmail.com}

\author{M\'onica A. Garc\'ia-\~Nustes}
\affiliation{Instituto de F\'isica, Pontificia Universidad Cat\'olica de Valpara\'iso, Casilla 4059, Chile}

\author{Jorge A. Gonz\'alez}
\affiliation{Department of physics, Florida International University, Miami, Florida 33199, United States}

\author{Juan F. Mar\'in}
\email[]{juan.marin.m@mail.pucv.cl}
\altaffiliation{Now at Departamento de F\'isica, Universidad de Santiago de Chile, Usach, Santiago, Chile}
\affiliation{Instituto de F\'isica, Pontificia Universidad Cat\'olica de Valpara\'iso, Casilla 4059, Chile}

\author{Diego Teca-Wellmann} 
\affiliation{Departamento de F\'isica, Universidad T\'ecnica Federico Santa Mar\'ia, Casilla 110-V, Valpara\'iso, Chile}


\date{\today}

\begin{abstract}

We investigate analytically and numerically the stability of bubble-like fluxons in disk-shaped heterogeneous Josephson junctions. Using ring
solitons as a model of bubble fluxons in the two-dimensional sine-Gordon equation, we show that the insertion of coaxial dipole currents
prevents their collapse. We characterize the onset of instability by introducing a single parameter that couples the radius of the bubble
fluxon with the properties of the injected current. For different combination of parameters, we report the formation of stable oscillating
bubbles, the emergence of internal modes, and bubble breakup due to internal mode instability. We show that the critical germ
depends on the ratio between its radius and the steepness of the wall separating the different phases in the system. If the steepness of
the wall is increased (decreased), the critical radius decreases (increases). Our theoretical findings are in good agreement with numerical
simulations. We discuss applications in quantum information technologies.

\end{abstract}


\maketitle



\section{Introduction}
\label{Sec:Introduction}

In the last three decades, a vast research area has been developed around the sine-Gordon (sG) system due to its many physical realizations
\cite{CuevasMaraver2014}. The most significant ones are the fluxon dynamics in Josephson junctions (JJ's) \cite{Ustinov1992,
Ustinov1998, Gorria2004, Lin2008}, spin waves in magnetic materials \cite{Clerc2008, BerriosCaro2016}, self-induced transparency in
nonlinear optics \cite{Blaauboer2000, Leblond2009}, and the Frenkel-Kontorova model of dislocations in solid state physics \cite{Geniet2002,
Braun2004, AlfaroBittner2017}. These applications are of great importance for both fundamental and technological motivations.
A flux quantum in a JJ can be controlled by bias currents \cite{Ustinov1998}, created or perturbed by the insertion of dipole
currents \cite{Ustinov2002, Malomed2004, Menditto2018}, pinned by built-in heterogeneities \cite{Aslamazov1984, Gurovich1987, Fehrenbacher1992}
or manipulated through shape engineering \cite{Gorria2004, Gulevich2006, Gulevich2007, Monaco2016}. Recent advances
in the design of wave parametric amplifiers \cite{OBrien2014, Macklin2015}, quantum information processing
protocols \cite{Macklin2015, Fujii2007, Kafri2016, Albarran2018} and the controlled manipulation of heat currents in superconducting devices
\cite{Guarcello2018, Guarcello2018-1, Guarcello2018-2, Guarcello2018-3} are very promising applications of this system.

A fluxon in a JJ is a physical realization of a sG soliton \cite{Peyrard2004, CuevasMaraver2014}. When a localized external dipole current
perturbs a fluxon, the internal modes of the fluxon can be locally destabilized to produce the insertion of a local fluxon-antifluxon pair. This
phenomenon is well understood in one-dimensional sG systems \cite{Gonzalez2002, Gonzalez2003}. In the two-dimensional case, the
authors have recently reported the formation of a flux line closed in a loop with a radially-symmetric bubble-like structure: a bubble fluxon
\cite{GarciaNustes2017}. The dynamics and stability of topologically equivalent structures in Klein Gordon (KG) systems have gained
significant attention in the literature, ranging from condensed matter to cosmology \cite{Geicke1984, Christiansen1997, Gorria2004, Ahmad2010,
Kevrekidis2018, Marin2018, Gonzalez2018, Kamranian2017, Johnson2012, Giblin2010}. Thus, theoretical studies on the formation, stability, and dynamics of bubble-like
structures are relevant not only for heterogeneous sG systems but also for more general systems.

In this article, we give an analytical and numerical study of the stability of these bubble-like structures by the application of external
forces. We demonstrate that the insertion of coaxial dipole currents may stabilize such structures. We provide a theoretical description of the
response of bubble fluxons to the coaxial dipole current intensity, finding a stabilization domain. Furthermore,
we show that internal mode dynamics generates remarkable phenomena, such as the formation of oscillating states, bubble instability and bubble
breakup. The article is organized as follows. In section \ref{Sec:SolitonJosephson}, we present the system and the model of bubble fluxons in
JJ's. In section \ref{Sec:ExternalForce}, we show analytically and numerically that the insertion of coaxial dipole currents in the junction
stabilizes bubble fluxons. In section \ref{Sec:Stability} we perform the linear stability analysis of the bubble solutions. Finally, we give
our conclusions and final remarks in section \ref{Sec:Conclusions}.

\section{Soliton bubbles in Josephson junctions}
\label{Sec:SolitonJosephson}

JJ's are built as two superconductors separated by a thin dielectric layer \cite{Barone1982}. A superconducting Josephson current composed by
tunneling Cooper pairs crosses the junction. This produces a jump on the phase $\phi$ of the wave function of the superconducting electrons
across the JJ. Such tunneling current may create a loop that continuously goes from one superconductor to the other, forming the so-called
Josephson vortex. These loops of current are represented by the well-known sG kink and antikink solutions, corresponding to a
$2\pi$-twist of superconducting phase $\phi$  \cite{Barone1982, Peyrard2004}. A Josephson vortex induces a local magnetic flux
trapped in a JJ. Each sG kink carries a quantum of magnetic flux $\Phi_o:=h/2e$, where $h$ is the Planck's constant. This is the so-called
fluxon. Kink solutions are associated with fluxons, whereas antikink solutions are associated with antifluxons. Each direction of the loop of
current is associated with the opposite signs of $\sin\phi(x)$ in different portions of the kink/antikink. Therefore, the center of the
kink/antikink corresponds to the position of the fluxon/antifluxon.

The propagation of fluxons in disk-shaped JJ's is governed by the following two-dimensional sG equation \cite{Barone1982, Peyrard2004}
\begin{equation}
 \label{Eq01}
 \partial_{tt}\phi-\nabla^2\phi+\gamma\partial_t\phi+\sin\phi=F(\mathbf{r}),
\end{equation}
where $\phi=\phi(\mathbf{r},t)$ is the quantum mechanical phase difference of the superconductors, $\mathbf{r}=(r,\theta)$ is the vector position in polar coordinates,
and $\nabla^2:=\partial_{rr}+r^{-1}\partial_{r}+r^{-2}\partial_{\theta\theta}$ is the Laplace operator. The third term of the left-hand side of
Eq.~\eqref{Eq01}, which plays the role of dissipation, represents the normal component of the tunnel current \cite{Malomed2014}. The
space-dependent force $F(\mathbf{r})$ represent non-uniform external perturbations introduced in the junction, such as dipole-current devices
\cite{Malomed1991, Malomed2004, Menditto2018}.

\begin{figure}
 \begin{center}
  \scalebox{0.35}{\includegraphics{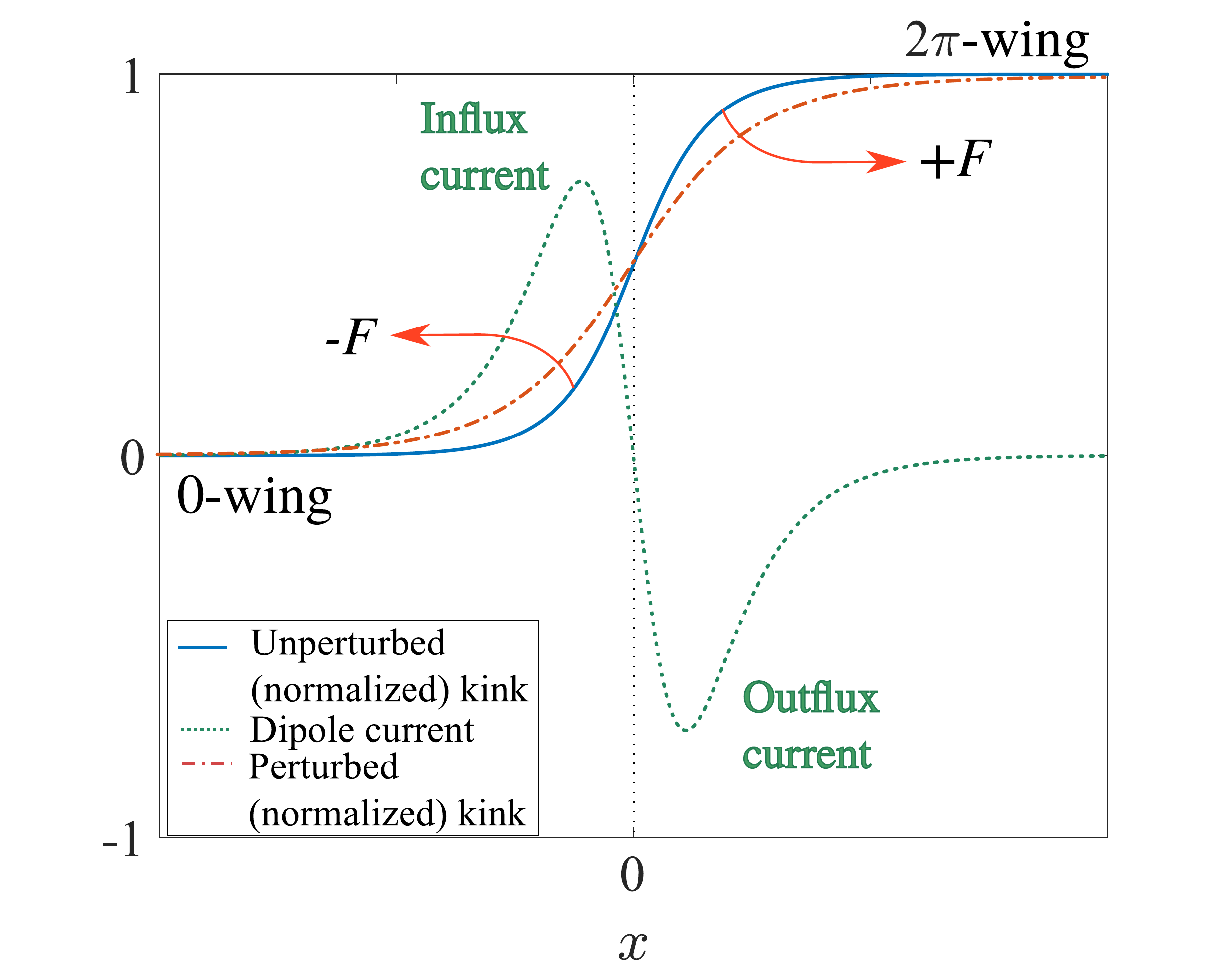}}
  \caption{(Color online) Dynamics of a fluxon under the influence of a current dipole device in a long (one-dimensional) Josephson
  junction. The kink profile of the fluxon (normalized by $2\pi$) is depicted in (blue) solid line. Under the influence of the current
  dipole, depicted in (green) dashed line, the fluxon is perturbed by counter-directed forces on its wings, producing a change on its 
  width. The perturbed fluxon profile (normalized by $2\pi$) is depicted in (red) dash-dotted line. \label{fig00}}
 \end{center}
\end{figure}

To investigate the stability of soliton bubbles in JJ's, we construct bubble-like profiles in the following manner. Circular ring soliton
solutions of Eq.~\eqref{Eq01} are given by $\phi_R(r)=4\arctan\exp[-(r-r_o)]$, where $r_o$ denotes the radius of the soliton
\cite{Christiansen1978, Christiansen1979, Christiansen1981}. In the case of a long one-dimensional JJ, it is known that a current dipole
device may perturb the shape and the width of a fluxon \cite{Gonzalez2002, Gonzalez2003, Gonzalez2006} (see Fig.~\ref{fig00}). A local
region of outflux (influx) current exerts a positive (negative) force in the $2\pi$-wing ($0$-wing) of the fluxon. The soliton is being
stretched due to these forces acting on its wings in opposite directions, changing its width. The perturbed kink in its steady state is
less steep than the unperturbed kink due to such stretching, as depicted in Fig.~\ref{fig00}. For sufficiently large values of the
influx/outflux current, the fluxon can be destroyed by the stretching. This is associated with an internal (shape) mode instability, a phenomenon
where the kink profile is no longer stable \cite{Gonzalez2002, Gonzalez2003}.

\begin{figure*}
 \begin{center}
\scalebox{0.2}{\includegraphics{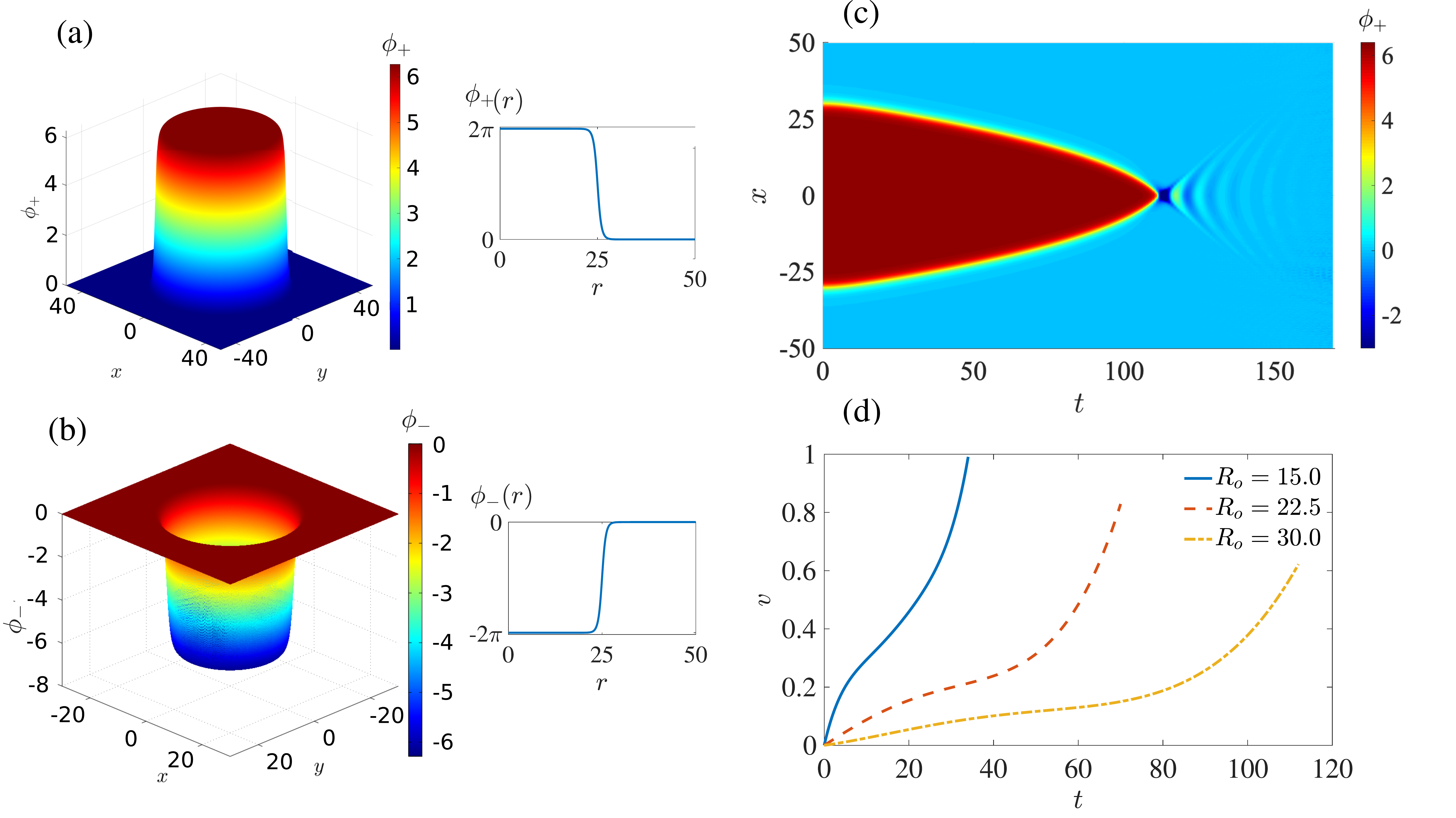}}
 \caption{(Color online) \textbf{(a)} Positive and \textbf{(b)} negative bubble-like fluxons. The insets show the profile of each structure
 as a function of the radial coordinate $r$. \textbf{(c)} Collapse of a positive bubble in the absence of external perturbations.
 Spatiotemporal dynamics for $r_o=30.0$, $\gamma=0.01$ and $B=1.6$. \textbf{(d)} Speed of the wall of bubbles as a function of time during
 their collapse. The initial radius of the bubble is indicated for each case.
 \label{fig01}}
 \end{center}
\end{figure*}

In the two-dimensional case, two-dimensional dipole currents may produce similar effects in the walls of a ring soliton. To take into
account the change in the width of the wall of a ring soliton under the influence of some external force, we introduce a
shape parameter $B$ according to $\phi_R(r;B)=4\arctan\exp[-B(r-r_o)]$. Moreover, following Ref.~\cite{GarciaNustes2017}, bubble-like
fluxons must have a kink/antikink profile at the walls and a well-defined derivative in the limit $r_o\to0$. To obtain bubble-like
profiles satisfying such conditions, we define the positive and negative bubbles as $\phi_+:=\phi_n-\phi_p$ and $\phi_-:=\phi_p-\phi_n$, respectively,
where $\phi_p(r):=4\arctan\exp[B(r-r_o)]$ and $\phi_n(r):=4\arctan\exp[B(r+r_o)]$. In Figs.~\ref{fig01}(a) and~\ref{fig01}(b) we depict these
solutions, which have a bubble-like shape indeed. It follows that the bubble solutions $\phi_{\pm}$ can be written in a more compact form as
\begin{eqnarray}
\label{Eq02}
 \phi_{\pm}(\mathbf{r})=4\arctan[\pm A\mbox{sech}(Br)],
\end{eqnarray}
where $A:=\sinh(Br_o)$ is a single real parameter that couples the two relevant parameters of our
system, namely $B$ and $r_o$.

Figure \ref{fig01}(c) shows the spatiotemporal evolution of the $x$-profile of $\phi_+$ at $y=0$. This has been obtained from the numerical
simulation of Eq.~\eqref{Eq01} with homogeneous Neuman boundary conditions, for $\gamma=0.01$ and $F=0$. For the Laplace operator, we have
used finite differences of second order of accuracy with steps $dx=dy=0.1$. The time integration was performed using a fourth order
Runge-Kutta scheme with step $dt=0.001$. The negative bubble profile $\phi_-$ collapses in a similar way. We conclude from these simulations
that bubble fluxons associated with these profiles are unstable structures. They collapse towards its center
under no external perturbations, and the structure is annihilated. This collapse is due to the kink-antikink attractive interactions between
the bubble walls and can be associated with the well-known return effect \cite{Christiansen1978}.

Let us consider the dynamics of a kink (line) soliton $\phi_{\ell}(x,y)=4\arctan[\exp(x)]$ in Eq.~\eqref{Eq01}, under the influence of a
constant and uniform external force $F(\mathbf{r})=F_o$. If $F_o<0$, the center-of-mass of the soliton will be accelerated in the positive
$x$-direction until its velocity saturates to a terminal velocity. Under small damping, such velocity will be near to the limiting value of
$c=1$. However, in the case shown in Fig.~\ref{fig01}(c), the external force is zero and the bubble wall is accelerated due to kink-antikink interactions. The strength of such interaction depends
on the distance between the walls. Thus, the walls of the bubble are under the action of a non-homogeneous force in space that enhances the
collapse of the bubble.

In Fig.~\ref{fig01}(d) we show the speed of the bubble-walls as a function of time for different bubbles with different initial radius.
These have been obtained tracking the position of the wall in time, and computing the time derivative of the highest
order well-conditioned time polynomial fitting the data. Initially, the bubble walls accelerate almost uniformly and tends to constant
terminal speed under the influence of small dissipation. However, the kink-antikink attraction becomes important as the radius of the bubble becomes
small. Such attraction produces a final burst in the speed just before the collapse, as noticed in Fig.~\ref{fig01}(d). The shorter
is $r_o$, the faster the collapse occurs and the greater is the speed just before the collapse.
We notice that bubbles with smaller radius collapse with higher accelerations. This
observation can be justified as follows. Let $\mathcal{A}:= F(r_o)/m(r_o)$ be the initial acceleration of the bubble wall, where
$m(r_o)=2\pi\,r_o\mathcal{M}$ is the inertial mass of the soliton bubble and $F=-dU(r_o)/dr_o$. The energy of the bubble wall is $U$ and
$\mathcal{M}$ is the mass of the bubble wall per unit length. If $\mathcal{U}$ is the energy of the bubble wall per unit length, then
$U=2\pi\,r_o\,\mathcal{U}$ and $F=-2\pi\mathcal{U}$, which is $r_o$ independent. It follows directly that
$a=-\mathcal{U}/(\mathcal{M}r_o)\propto1/r_o$. Such decay law holds if $r_o$ is large enough so that kink-antikink interactions between
the bubble walls are negligible. Thus, we expect that the scaling law would be different just before the collapse. The
acceleration exerted on the walls is not uniform in space, becoming larger as the radius becomes smaller. This produces a final burst in the
velocity profiles, instead of the typical saturation profile observed for homogeneously driven kinks under small damping.

\section{Controlling bubbles with external perturbations}
\label{Sec:ExternalForce}

\begin{figure*}
  \scalebox{0.3}{\includegraphics{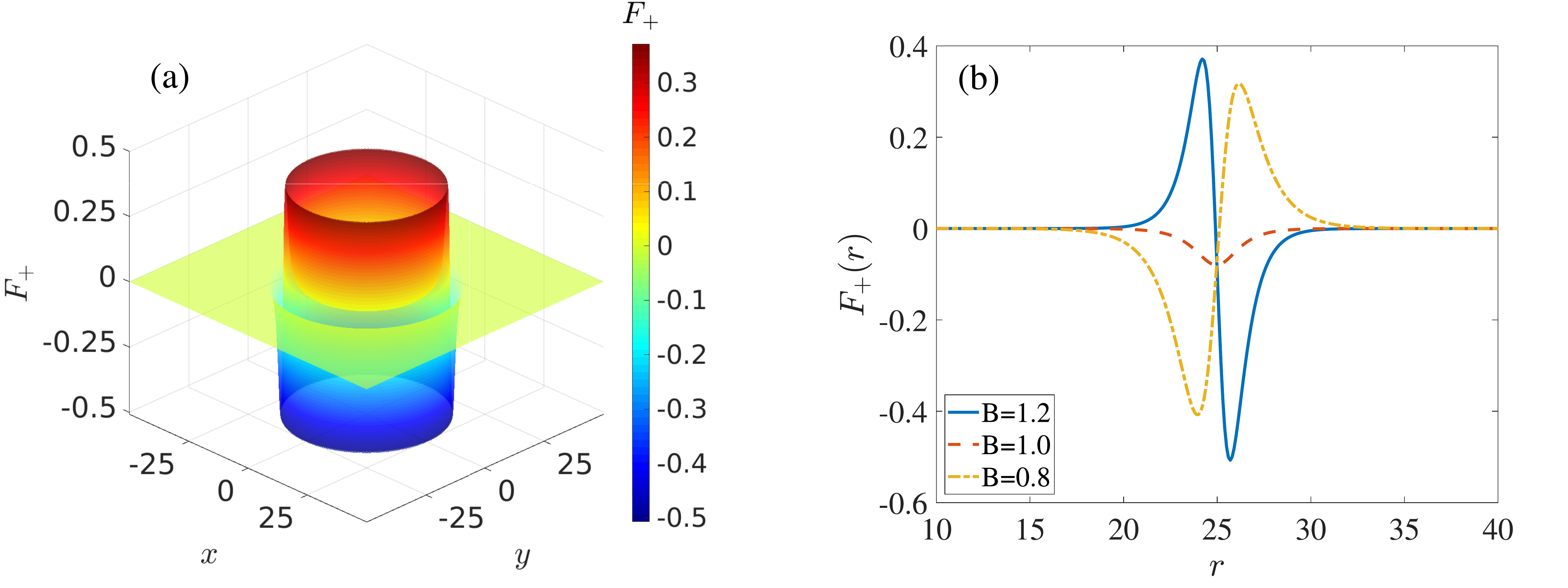}}
    \caption{(Color online) \textbf{(a)} The ringlike force of Eq.~\eqref{Eq03} for $B=1.2$, modeling a coaxial current dipole with a radius $r_o=25.0$
    inserted into the Josephson junction. \textbf{(b)} Profile of $F_{+}$ as a function of $r$ for $r_o=25.0$ and different values of
    $B$.}
  \label{figureforce}
\end{figure*}

For quantum information technologies, it is of crucial importance the control, the insertion, stability, and transport of fluxons. Thus, a
natural question is whether we can control the bubble structures of the previous section by the application of some external perturbation.
Indeed, we have obtained that the insertion of a coaxial current dipole device in the junction can stabilize such structures.

Fluxon dynamics in JJ's under the influence of a current dipole device is modeled through an additive-- external --force in the sG equation
\cite{Ustinov2002, Malomed2004}. Here, to calculate the family of forces that can stabilize these bubble fluxons, we solve an inverse problem
similarly to previous works \cite{Gonzalez2006, Gonzalez1992, Gonzalez2002, Gonzalez2003}. Let us consider the bubble profile of
Eq.~\eqref{Eq02} as the initial condition. After substitution in the sG Equation \eqref{Eq01}, we obtain that $\phi_{\pm}$ is
an exact solution if the external force is given by 
\begin{widetext}
\begin{eqnarray}
 \label{Eq03}
F_{\pm}(\mathbf{r})=\left[\pm2(B^2-1)\alpha(r)+\frac{2B}{r}\beta(r)\tanh(Br)+\Upsilon\left(\alpha(r)-\beta(r)\right)\right]\mbox{sech}(Br),\\
\alpha(r):=-2A\frac{1-A^2\mbox{sech}^2(Br)}{\left[1+A^2\mbox{sech}^2(Br)\right]^2},\quad
\beta(r):=\frac{-2A}{1+A^2\mbox{sech}^2(Br)},
\end{eqnarray}
\end{widetext}
where $\Upsilon:=2B^2/A^2$. Figure \ref{figureforce} shows $F_{+}$ for the given values of parameters.
It is important to remark that as $r\to0$, the second term in the square brackets of Eq.\eqref{Eq03}
goes to zero since the factor $\beta(r)\tanh(Br)$
decays to zero exponentially in that limit.
Regions with $F_{\pm}<0$ ($F_{\pm}>0$) corresponds to an area in the junction with an
influx (outflux) of current. A coaxial dipole current is defined by an influx and
outflux zone separated a
distance $D$. Given the current distributions, such zones have
a finite decay width. This latter is physically consistent with an actual dipole
current in experimental setups, where point-like
dipoles are not realistic. If the distance $D$ is of the order of the decay width, the dipole is no longer defined. The dipole existence
range is determined by the interplay of both $B$ and $r_0$.
In Fig.~\ref{figureforce}(b) we show that $F_+$ represents a coaxial dipole current for
$r_0=25.0$, $B=1.2$ and $B=0.8$. For $B=1$, the distance $D$ is smaller than the decay
width and, therefore, the force $F_+$ cannot be regarded as a dipole current.
If the force has a dipole profile, such a pair of influx/outflux current generates a local magnetic field that affects
the dynamics of fluxons in the junction. Indeed, from a purely geometrical point of view, this ring-like force can be regarded as a solid
of revolution formed by the dipole currents considered in Refs.~\cite{GarciaNustes2017, Malomed2004, Ustinov2002}.
The total current $I\propto\int dV\,F_{\pm}(r)$ from the dipole is always negative. Therefore, the dipole is injecting current
into the junction, providing energy that can be used to stabilize fluxon bubbles.
The axial symmetry of the force allows the manipulation of rotationally symmetric localized structures, such as bubble-solitons. Notice that
parameter $B$ is associated with both the intensity of the current and the spatial extension of the injection area.

\begin{figure*}
\scalebox{0.265}{\includegraphics{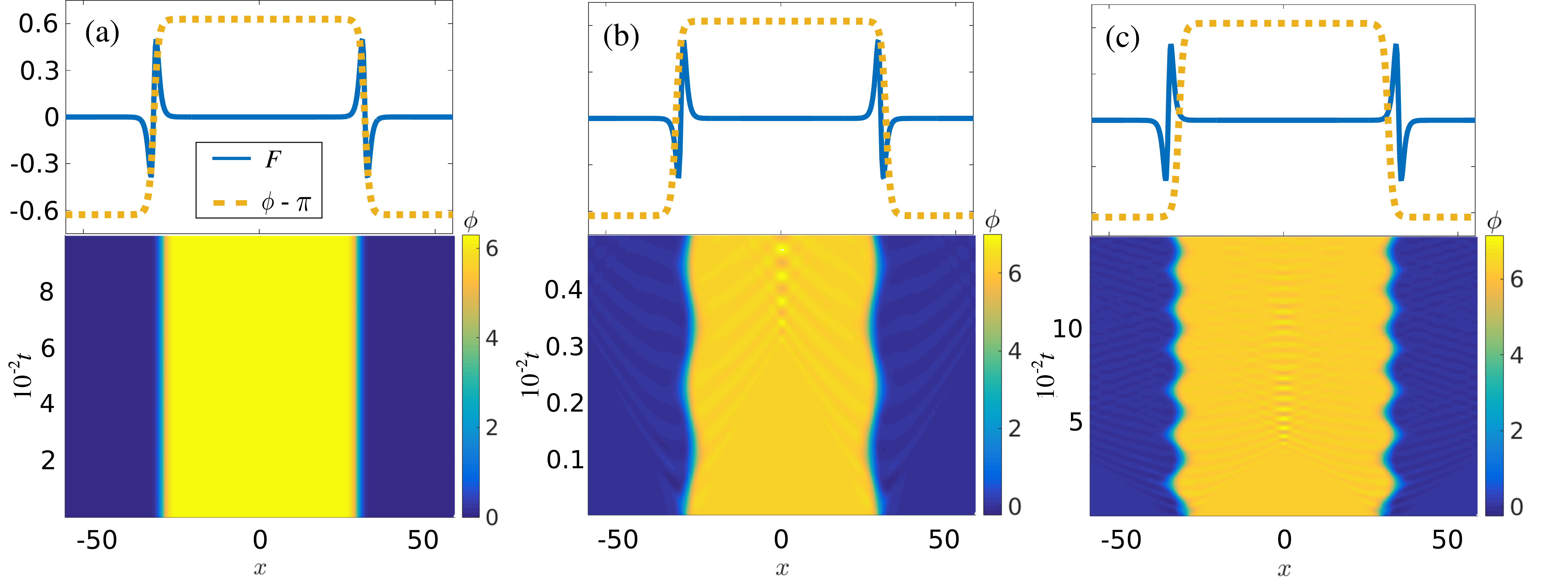}}
\caption{(Color online) Spatiotemporal evolution of bubble fluxons with an initial radius $r_o$ under the action of a coaxial dipole current with a radius $R$,
for $\gamma=0.01$ and $B=1.2$. \textbf{(a)} A stable positive bubble for $r_o=R=30.0$. \textbf{(b)} An oscillating state for $r_o=30>R=28.5$.
The area of the bubble performs persistent oscillations around the equilibrium radius $R$. \textbf{(c)} An oscillating state for
$r_o=30.0<R=33.0$. The upper insets show the profile at $y=0$ of the initial bubble and the ring-like force for each case \cite{SuppMat}.
\label{fig02}}
\end{figure*}

Figure \ref{fig02}(a) shows a positive bubble perfectly stabilized by $F_+$. This is the equilibrium configuration of
the system, where both the bubble and the force has the same radius (see inset with the profiles of $\phi_+$ and $F_+$ at $y=0$). Moreover,
we can also consider coaxial forces with a different radius than the initial bubble. Numerical simulations show that for $B=1.2$, the force stabilizes bubbles if the difference between the radii is at most $10\%$ of the bigger radius. That is, if the initial radius of the
bubble is smaller by more than a $10\%$ of the radius of the force, then the coaxial dipole current is too far from the walls of the
bubble to prevent its collapse. The bubble collapses almost as it were unperturbed. If on the other hand, the initial radius of the bubble is
bigger by more than a $10\%$ of the radius of the force, then the bubble begins to shrink for enough time to gain relatively high
speed. The coaxial dipole cannot capture the wall of the bubble because it passes very fast, and the bubble collapse.

In any other circumstance, the wall of the bubble does not have enough time to gain too much speed, and the coaxial dipole traps the bubble
preventing its collapse. In Fig.~\ref{fig02}(b), we show the formation of an oscillating bubble solution. The initial condition is a
fluxon bubble whose radius is only $5\%$ bigger than the radius of the force. The bubble shrinks and is captured by the dipole,
making sustained oscillations in time around the equilibrium configuration. In Fig.~\ref{fig02}(c) we show the formation of a similar oscillating
state from a bubble whose radius is nearly $10\%$ smaller than the radius of the force. In this case, the bubble is expanded due to
an attractive interaction with the coaxial dipole, making again sustained oscillations around the equilibrium configuration. These stable and
oscillating structures emerge from the compensation of two counteractive forces: one associated with the return effect, and another associated
to the coaxial dipole current. We will return to this point later in this article.

\section{Stability analysis of bubble fluxons}
\label{Sec:Stability}

In this section, we perform a linear stability analysis of the bubble solutions of Eq.~\eqref{Eq02} under the action of
the ring-like force of equation \eqref{Eq03}. For that purpose, we write the perturbed sG equation \eqref{Eq01} as
\begin{equation}
\label{Eq05}
 \partial_{tt}\phi-\nabla^2\phi+\gamma\partial_t\phi-G(\phi)=F(\mathbf{r}),
\end{equation}
where $G(\phi):=-dU/d\phi$, and $U(\phi):=1-\cos\phi$. To investigate the stability of the structure $\phi_{\pm}$, let us consider a
small-amplitude perturbation $\chi$ around such solution, i.e.
\begin{eqnarray}
 \label{Eq06a}
 \phi(\mathbf{r},t)&=& \phi_{\pm}(\mathbf{r})+\chi(\mathbf{r},t)\\
 \label{Eq06b}
 \chi(\mathbf{r},t)&:=&f(\mathbf{r})e^{\lambda t}\quad,\quad |\chi|\ll|\phi_{\pm}|\forall(\mathbf{r},t).
\end{eqnarray}

Expanding $G(\phi)$ with $\phi\sim\phi_{\pm}$ neglecting terms of order $\mathcal{O}(|\chi|^2)$, and after substitution of \eqref{Eq06a} and
\eqref{Eq06b} in the sG equation \eqref{Eq01} with $F(\mathbf{r})=F_{\pm}(\mathbf{r})$, we obtain for $f(\mathbf{r})$ the following  eigenvalue problem
\begin{equation}
 \label{Eq07}
 -\mathbf{\nabla}^2f+V_{\pm}(r)f=\Gamma f,
\end{equation}
where $\Gamma:=-\lambda(\lambda+\gamma)-1$ and $V_{\pm}(r):=\cos\phi_{\pm}(\mathbf{r})-1$. Equation \eqref{Eq07} is equivalent to the
time-independent Schr\"odinger equation. The potential $V_{\pm}(r):=V(r)$ is the same for both structures, the positive and negative bubbles,
and is given by
\begin{equation}
 \label{Eq08}
 V(r)=-\frac{8A^2\mbox{sech}^2(Br)}{\left[1+A^2\mbox{sech}^2(Br)\right]^2}.
\end{equation}

\begin{figure}
    \scalebox{0.3}{\includegraphics{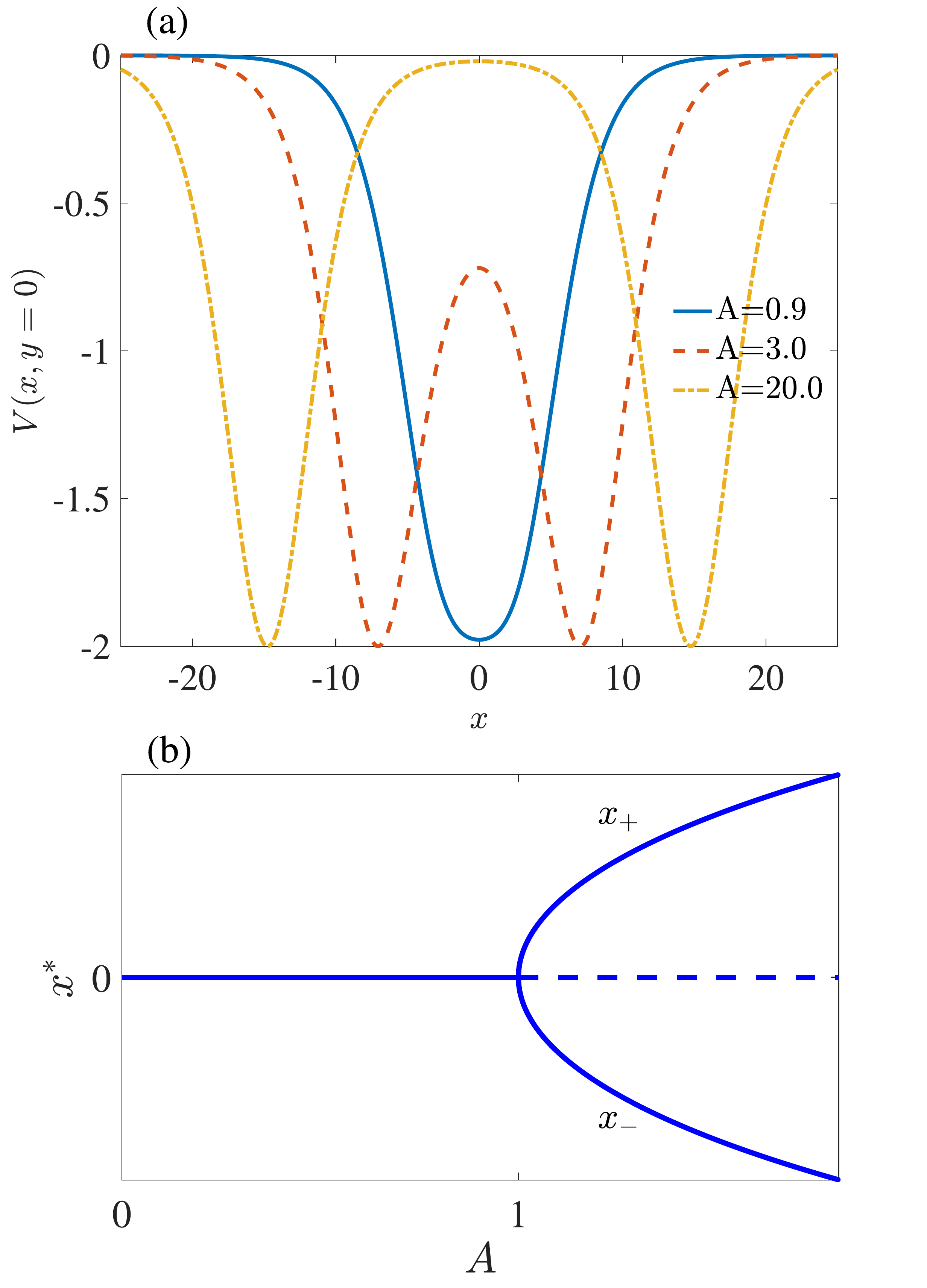}}
  \caption{(Color online) \textbf{(a)} Potential of Eq.~\eqref{Eq08} for $B=0.25$ and the indicated values of $A$. At $A=1$, the system bifurcates
  from a single-well to a double-well potential. \textbf{(b)} Bifurcation diagram of $V(x,y=0)$, showing the position of the equilibrium (fixed
  points) at the origin, $x_+$ and $x_-$ as a function of the bifurcation parameter $A$. A Pitchfork bifurcation occurs at $A=A_c=1$. Solid lines denote
  stable fixed points, whereas dashed lines denote unstable fixed points.}
 \label{potential}
\end{figure}

From the potential \eqref{Eq08}, we can  identify two delimited regions of parameter $A$ for which the system exhibit different qualitative
behaviors. Figure~\ref{potential}.a shows the profile at $y=0$ of the potential~\eqref{Eq08} for different values of $A$. For $A\leq1$,
Eq.~\eqref{Eq08} at $y=0$ is a hyperbolic potential well, while for $A>1$ is a hyperbolic double-well potential, as shown
in Fig.~\ref{potential}.a. Indeed, for $A<1$ the potential \eqref{Eq08} has only one real and stable equilibrium point at the origin. Above
the critical value $A=A_c:=1$, the equilibrium point at the origin turns unstable, and two new real stable points appears at
$x=x_{\pm}:=\pm B^{-1}\mbox{acosh}(A)$. Thus, at $A=A_c$ the profile of potential $V$ passes from a single-well to a double-well 
structure through a Pitchfork bifurcation, as shown in Fig.~\ref{potential}.b. The system has very different dynamics in each region, which
are separated by the bifurcation point $A=A_c$. In the following, we investigate the dynamics of the system in each region.

\begin{figure}
\scalebox{0.3}{\includegraphics{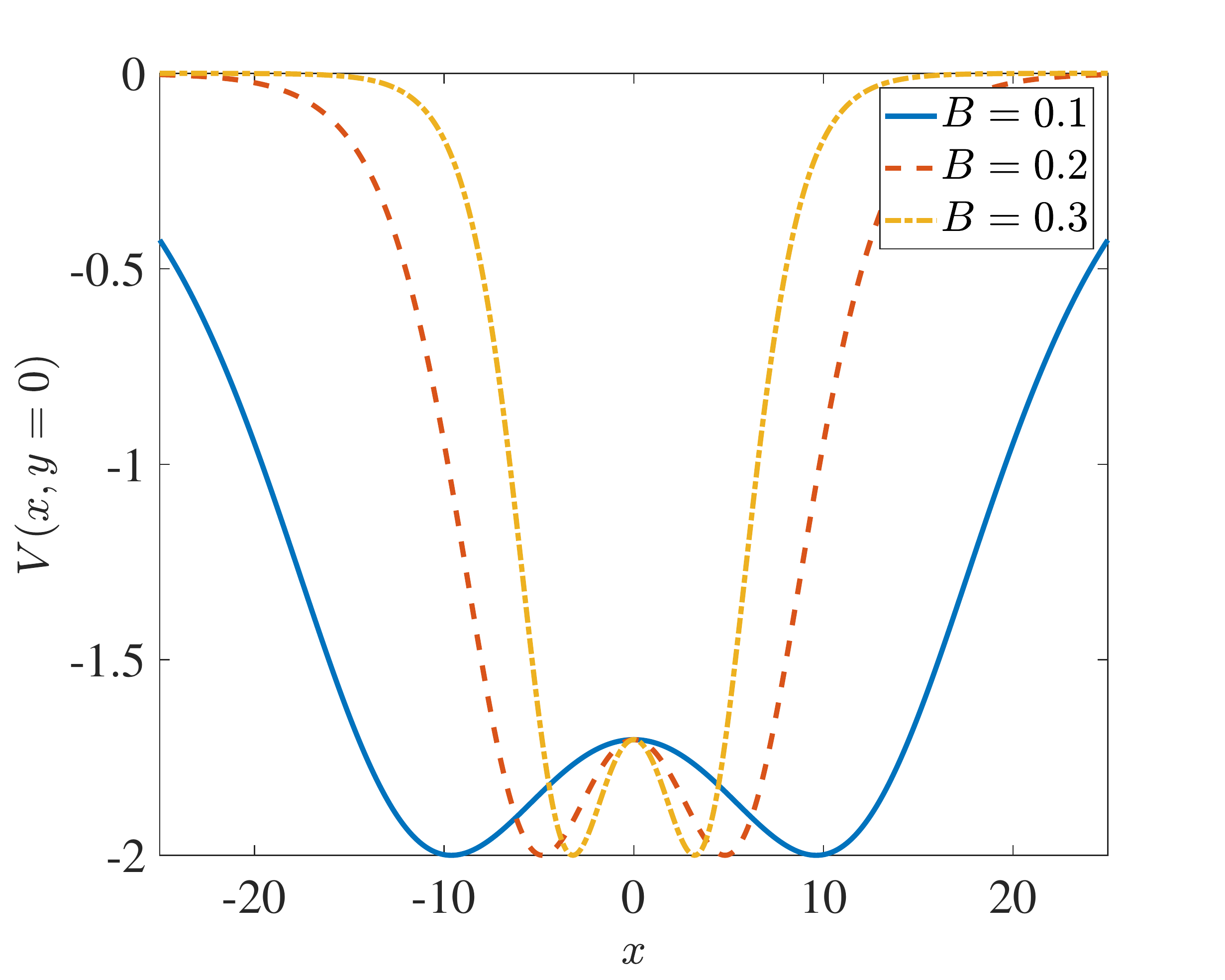}}
\caption{(Color online) The double-well potential of Eq.~\eqref{Eq08} at $y=0$ for $A=1.5$ and the given values of $B$. The walls of the
double-well become steeper as $B$ increases.\label{fig04}}
\end{figure}

\subsection{The double-well region ($A>1$): Stable bubbles, oscillating states, generation of internal modes and bubble insertion}

In the case $A>1$, the potential of Eq.\eqref{Eq08} has a double-well structure, as depicted in Fig.~\ref{fig04} for different values of $B$.
We notice that if $A\to\infty$, then $V(0)\to0$ and $x_{\pm}\to\infty$, which means that the separation of the minima of the double-well
increases indefinitely as $A\to\infty$. We notice this
behavior in Fig.~\ref{potential}. Therefore, in the limiting case $A\gg1$, the double-well potential \eqref{Eq08} can be regarded as two
independent single wells very far from each other. Let us define $\xi:=\mbox{sech}(Br)$ and $\xi_{\pm}:=\mbox{sech}(Bx_{\pm})$.
After a Taylor expansion of the potential with $\xi\sim\xi_{\pm}$, and noticing that $\xi_{\pm}\to0$ as $A\to\infty$, one obtains that the
potential behaves locally for $r\sim x_+$ as
\begin{equation}
 \label{Eq09}
V(r)\simeq-2\mbox{sech}^2[B(r-x_{+})],
\end{equation}
which is the so-called modified P\"oschl-Teller potential hole. Thus, away from the bifurcation point, the double-well potential of
Eq.~\eqref{Eq08} behaves as two effective P\"oschl-Teller potentials very far from each other.

Notice that from the condition $A\gg1$
follows $Br_o\gg1$, which means that the radius of the bubble is much greater
than $1/B$. In such limit the curvature effects are negligible
($\nabla^2\simeq\partial_{rr}$). Thus, the system can be regarded as
quasi-one-dimensional, obtaining a good estimate of the bound states of the eigenvalue problem \eqref{Eq07} by solving the
Schr\"odinger equation at each well separately. Indeed, the
Schr\"odinger equation \eqref{Eq07} for the P\"oschl-Teller potential can be solved exactly
\cite{Flugge2012}, and has appeared previously in the literature for the stability analysis of many nonlinear structures, such as
one-dimensional sine-Gordon kinks \cite{Holyst1991, Gonzalez2002, Gonzalez2003, Gonzalez2006, Gonzalez1992}.
The eigenfunctions determine the oscillations around the bubble solution. The scattering
states, corresponding to the continuous spectrum, are generally called phonon modes \cite{Peyrard1983}. Meanwhile, the soliton modes correspond to the bound states, whose
eigenvalues lie in the discrete spectrum \cite{Gonzalez2007} and are given by the formula
\begin{equation}
 \label{Eq10}
 \Gamma_n=B^2(\Lambda+2\Lambda n-n^2)-1,
\end{equation}
where $\Lambda(\Lambda+1)=2/B^2$, and $n=0,\,1,\,\ldots,\, [\Lambda]-1$, where $[\Lambda]$ is the integer part of $\Lambda$.

Notice that for $B^2=1$ we obtain $\Gamma_0=0$, and the system possesses translational invariance for $r$ sufficiently large. This zero-frequency
bound state is the Goldstone mode, which is associated with the translational mode. Here on, we will refer to this mode as the translational
mode of the bubble walls. If there exist other modes with $n>0$, they will be internal shape modes and will be responsible for the vibrations
of the walls.

\begin{figure}
 \begin{center}
  \scalebox{0.3}{\includegraphics{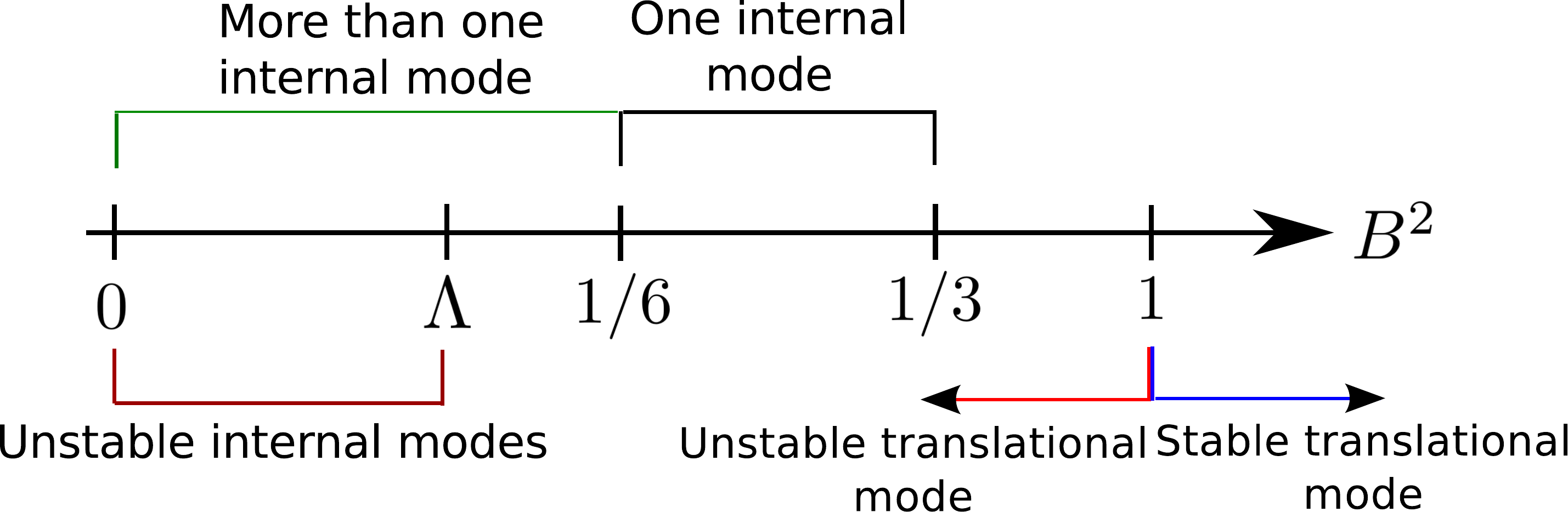}}
  \caption{Schematization of the results obtained from the linear stability analysis in the double-well region ($A\gg1$).}
  \label{FigBline}
 \end{center}
\end{figure}

We can now predict theoretically the response of fluxon bubbles to the coaxial dipole current for different values of the control parameter
$B$. Figure \ref{FigBline} summarizes the results. Such results are valid for both positive and negative bubble solutions.

\begin{figure*}[t]
 \begin{center}
\scalebox{0.27}{\includegraphics{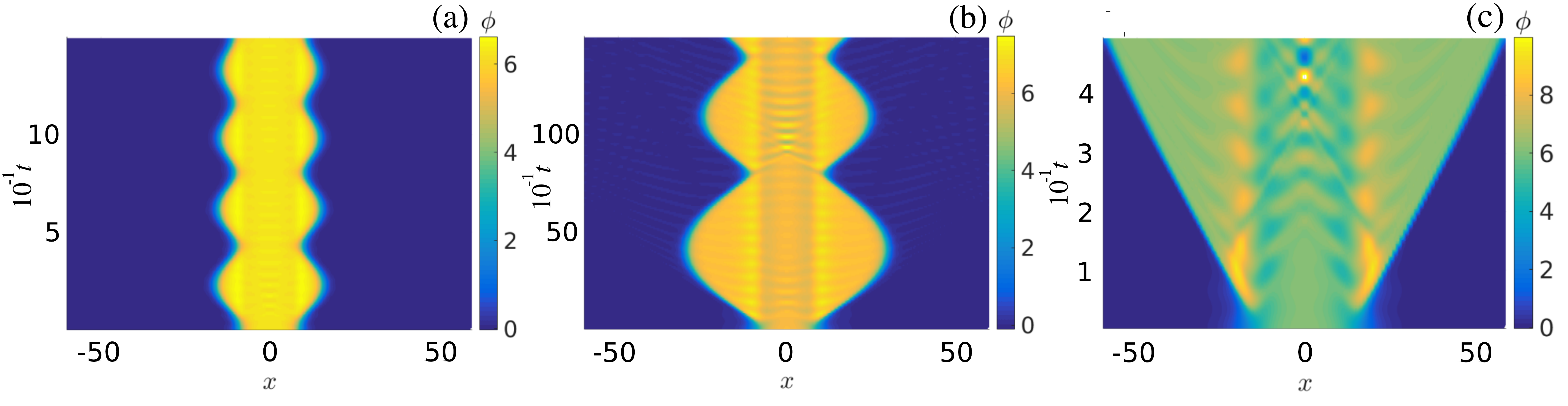}}
 \caption{(Color online) \textbf{(a)} A stable bubble due to the equilibrium of competing forces for $(B,\,r_o,\,r_f)=(0.8,\,8.7,\,8.3)$.
 \textbf{(b)} A stable bubble oscillating at a lower frequency for $(B,\,r_o,\,r_f)=(0.5,\, 8.7,\,8.3)$. The oscillations from a stable
 internal mode can be clearly appreciated. \textbf{(c)} An unstable bubble for $(B,\,r_o,\,r_f)=(0.26150,\, 16,\,15.3)$. The oscillations of
 the internal mode become more complex due to the emergence of more than one internal mode \cite{SuppMat}.}
 \label{fig07}
 \end{center}
\end{figure*}

\subsubsection{Stability of the translational mode}
\label{UnstableTrasnsmode}

If $B^2>1$, we obtain $\lambda_0<0$ and the translational mode is stable. In this case, the ring-like force stabilizes the soliton bubble to a
fixed radius equals to the radius of the force itself. Indeed, the combination of parameters of the numerical simulations showed in
Fig.~\ref{fig02} corresponds to this case, and the theory predicts the stability of the bubble correctly. In the simulation of
Fig.~\ref{fig02}(a), the core of the bubble wall is exactly at the stable equilibrium position, and the bubble is completely stationary.
If we displace the core of the wall from this equilibrium point slightly, then the wall oscillates around the stable equilibrium--see
figures \ref{fig02}(b) and \ref{fig02}(c).

If $1/3<B^2<1$, the translational mode becomes unstable and there are no internal modes, given that $[\Lambda]<2$. Under these conditions, the
equilibrium points $x_{\pm}$ are now unstable. Therefore, if the initial radius of the bubble is smaller than the radius of the ring-like force, 
the bubble will collapse. If on the contrary, the initial radius of the bubble is bigger than the radius of the force, the external force will
push the bubble wall away, and the bubble will expand. Eventually, the curvature force that tends to collapse the bubble compensates the
repulsion due to the ring-like force. The acceleration of the bubble expansion will be zero when one force equilibrates the other, and
eventually, the bubble wall will go backward.

The dynamical behavior observed in Fig.~\ref{fig07}(a) agrees with these theoretical findings. The initial condition is a positive
bubble, which has a radius $r_o$ slightly bigger than the radius $r_f$ of the coaxial dipole current. The
instability of the translational mode produces a motion of the bubble-wall away from the unstable fixed point located at the zero of the force.
This instability along with the return effect produce the oscillatory behavior of the bubble area, as expected.

 \begin{figure}
  \begin{center}
  \scalebox{0.35}{\includegraphics{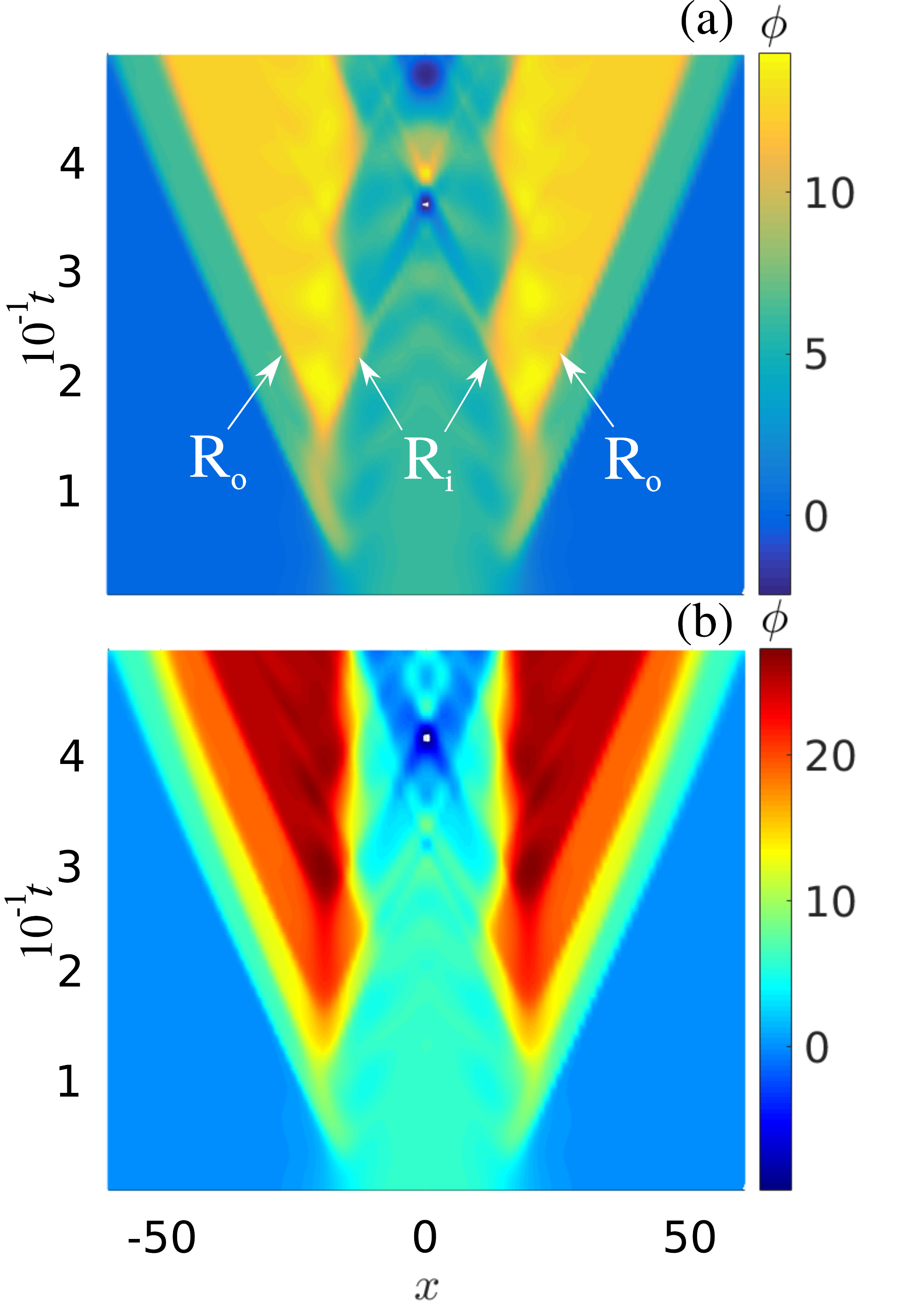}}
  \caption{(Color online) Bubble breakup obtained for \textbf{(a)} $B=0.261$, and \textbf{(b)} $B=0.255$. In both cases, $r_o=16$ and $r_f=15.3$
  \cite{SuppMat}}
  \label{fig08}
  \end{center}
 \end{figure}

\subsubsection{The existence of stable internal modes}

The repulsion from the dipole current becomes stronger, and the translational mode turns more unstable as $B$ decreases. Indeed, we notice in
the simulation of Fig~\ref{fig07}.(b) that the bubble reaches a bigger radius during the first burst than in Fig.~\ref{fig07}(a). The
oscillating states of Fig.~\ref{fig02} are similar to the oscillating state of Fig.~\ref{fig07}(a). However, in the latter, the bubble area
describes damped oscillations. The amplitude of the damped oscillations are more prominent, the frequency decreases,
and the decay of the oscillations is stronger as $B$ decreases.

Although the dynamics of the bubble in Fig.~\ref{fig07}(b) is similar as in Fig.~\ref{fig07}(a), we notice the presence of small oscillations
near the zero of the ring-like force in this last case. This is, indeed, evidence of the existence of a stable oscillatory--internal--mode in the system. Indeed,
if $1/6<B^2<1/3$, the translational mode is unstable and the first internal mode of the bubble will appear. In this case, we obtain
$\lambda_1<0$. Therefore, this internal mode is stable. The width of the bubble walls will be oscillating at a frequency equal to
$\omega=\sqrt{\Gamma_1}$. The numerical simulation of Fig.~\ref{fig07}(b) is in this range of parameters. Given that the translational mode of
the wall is still unstable in this case, the bubble area initially expands. The parameter $B$ is lower in this case than in the previous
simulation. Thus, the maximum of the ring-like force is bigger and the bubble expands.

In case $B^2<1/6$, more internal modes will appear, and the oscillations in the walls become complex. There is a
contribution of more than one frequency in this case. The simulation of Fig.~\ref{fig07}(c) illustrates this scenario. More than one stable internal modes are present in the system, and the kink-antikink attractive force is too weak compared to the repulsion from
the current dipole. The translational mode is highly unstable, and the bubble expands.

\subsubsection{Bubble breakup by internal mode instability}

  \begin{figure}
  \begin{center}
  \scalebox{0.5}{\includegraphics{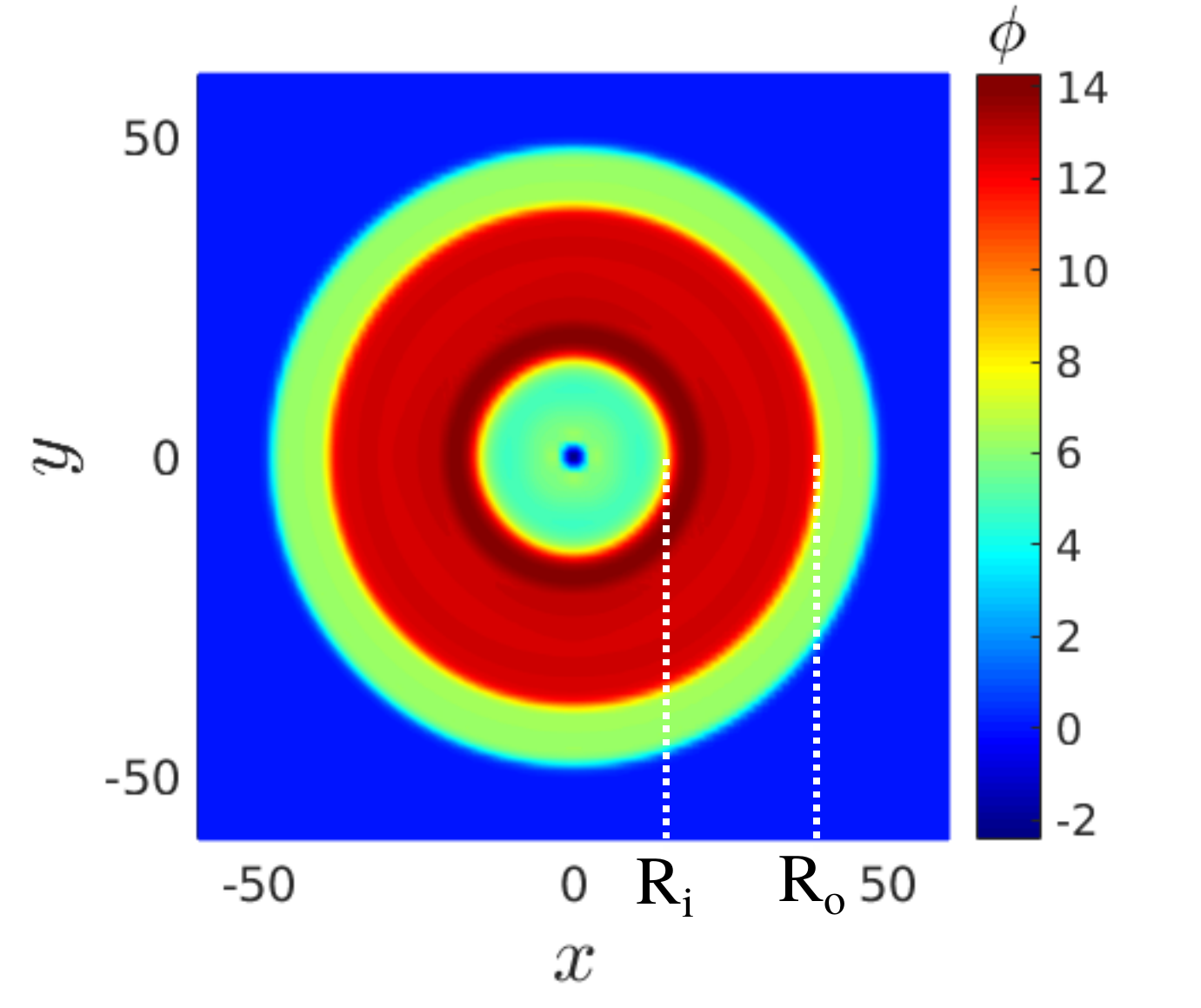}}
  \caption{(Color online) An expanding $2\pi-4\pi$ disk-like bubble generated inside the initial $0-2\pi$ bubble for $B=0.261$, $r_o=16$,
  $r_f=15.3$ and $t=0.725$ \cite{SuppMat}.}
  \label{fig08a}
  \end{center}
 \end{figure}
 
We have seen that for sufficiently small values of $B$, the translational mode of the bubble wall can become unstable. An important question
is whether the internal modes of the wall can also become unstable for sufficiently small values of $B$. Indeed if
$B^2<2/\Lambda^*(\Lambda^*+1)$, where $\Lambda^*=(5+\sqrt{17})/2$, the first internal mode becomes unstable! Figure \ref{fig08} shows the
bubble breakup. The initial bubble expands due to the instability
of the translational mode, but soon after this expansion, more structures are inserted in the system by the coaxial dipole current. This
insertion of new structures is due to the creation of a pair kink-antikink in the wall of the initial bubble.

Figure \ref{fig08}(a) shows that a new structure is inserted in the system around $t=0.17$ for $B=0.261$. The internal shape-mode
instability produces the breaking of the wall of the bubble, and it acquires a ladder-like profile that expands in time. This new structure is a
\emph{disk-shaped bubble}, whose inner and outer radius is denoted as  $R_i$ and $R_o$, respectively. The inner (outer)
wall of the inserted disk-shaped bubble is an antikink (kink), and therefore is trapped (ejected) by the coaxial force. Thus, $R_i$ performs
damping oscillations around the equilibrium point $r_f$ while $R_o$ grows in time, as we show in Fig.~\ref{fig08}(a). The 
generated disk-shaped bubbles are always expanding structures.

The number of inserted bubbles depends on the actual value of parameter $B$. Figure \ref{fig08}(b). shows that more bubbles are rapidly inserted
in the system for $B=0.255$. The generated traveling ladder has more steps for decreasing values of parameter $B$. Figure \ref{fig08a}
shows a typical configuration of the system after the initial breakup of the bubble, where an expanding disk-shaped bubble is appreciated. 

\subsection{The single well region ($A<1$): Unstable bubbles}
 
In the case $A<1$, Eq.~\eqref{Eq08} for $y=0$ is a potential well--as previously shown in Fig.~\ref{fig04}. Moreover, in the limiting case $A\ll1$
it reduces to
\begin{equation}
 \label{Poschl-Teller}
V(r)=-8A^2\mbox{sech}^2(Br),
\end{equation}
which is also a P\"oschl-Teller potential hole whose exact solutions are known. However, here we show that in this region of parameters the
force $F_{\pm}$ is no longer a coaxial dipole, and the system does not support stable bubble solutions.

\begin{figure}
  \begin{center}
  \scalebox{0.3}{\includegraphics{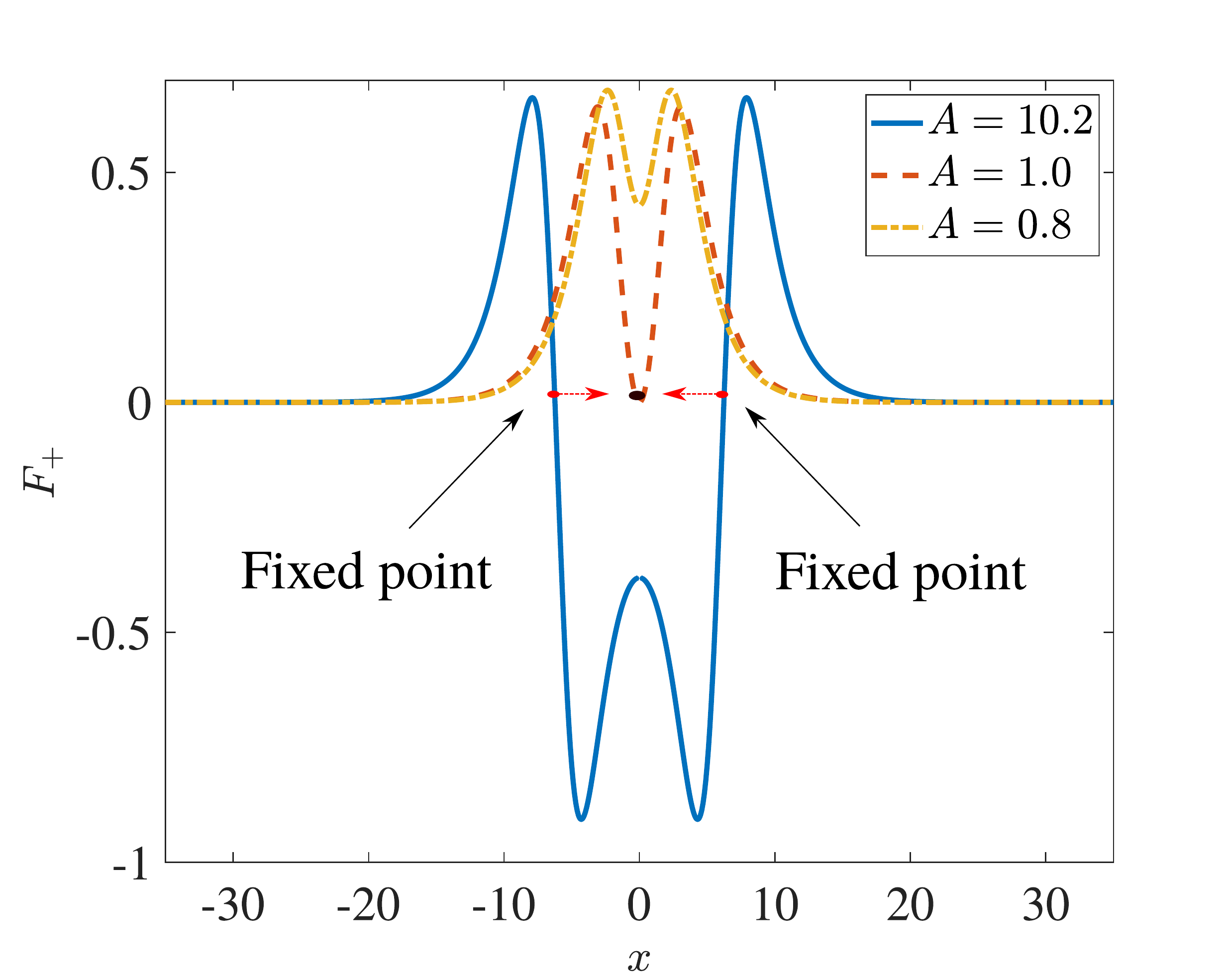}}
  \caption{(Color online) Profiles of the coaxial force $F_+$ for $B=0.5$, $y=0$, and the indicated values of $A$ as a function of $x$. For
  $A=10.2>1$ ($r_o\simeq6.0359$), the force has two fixed points for the position of the bubble walls. At $A=1$ ($r_o\simeq1.7627$), such stable points
  collide at the origin. For $A=0.8<1$ ($r_o=1.4653$), there are no fixed points, and the system does not support stable bubble solutions.
  }
  \label{fig09}
  \end{center}
 \end{figure}

In sG systems, as in more general KG systems, it is well known that the zeroes of an external force are fixed points for
kink and antikink solutions \cite{Gonzalez2002, Gonzalez2003}. Given that the walls of the bubble solutions $\phi_{\pm}$ can be regarded as a
pair kink-antikink in the $r$-domain, the zeroes of $F_{\pm}$ are fixed points that may stabilize the structure. Figure \ref{fig09} shows the
force as a function of $x$ at $y=0$ for the given values of $A$. For $A>1$, that is in the double-well regime, the force has two fixed points that
may contribute to the stabilization of bubble fluxons. If the fixed points are stable, a nearby bubble wall will be attracted by the fixed
points, and either a stationary or an oscillatory bubble will be observed. This is the case of the simulations of Fig.~\ref{fig02}. On the
contrary, if the fixed points are unstable, then a nearby bubble wall will be ejected from the fixed point, and one would observe one of the
structures reported in figures \ref{fig07} and \ref{fig08}.

Diminishing the value of $A$ towards $A_c:=1$, the system approaches the bifurcation point, and the zeroes of the force approach
each other. This is indicated in Fig.~\ref{fig09} by the direction of the red arrows. The minima of the associated double-well potential
have similar behavior as the fixed points of the walls as $A\to1$. At the bifurcation point, the two fixed points collide and
annihilate each other! At this point, the external force is no longer a coaxial dipole. For $A<1$ there are no fixed points, and the system does not support
stable bubble solutions anymore. Numerical simulations have confirmed our analysis. All bubbles in this region of parameters collapse.
 
\begin{figure}
  \begin{center}
  \scalebox{0.3}{\includegraphics{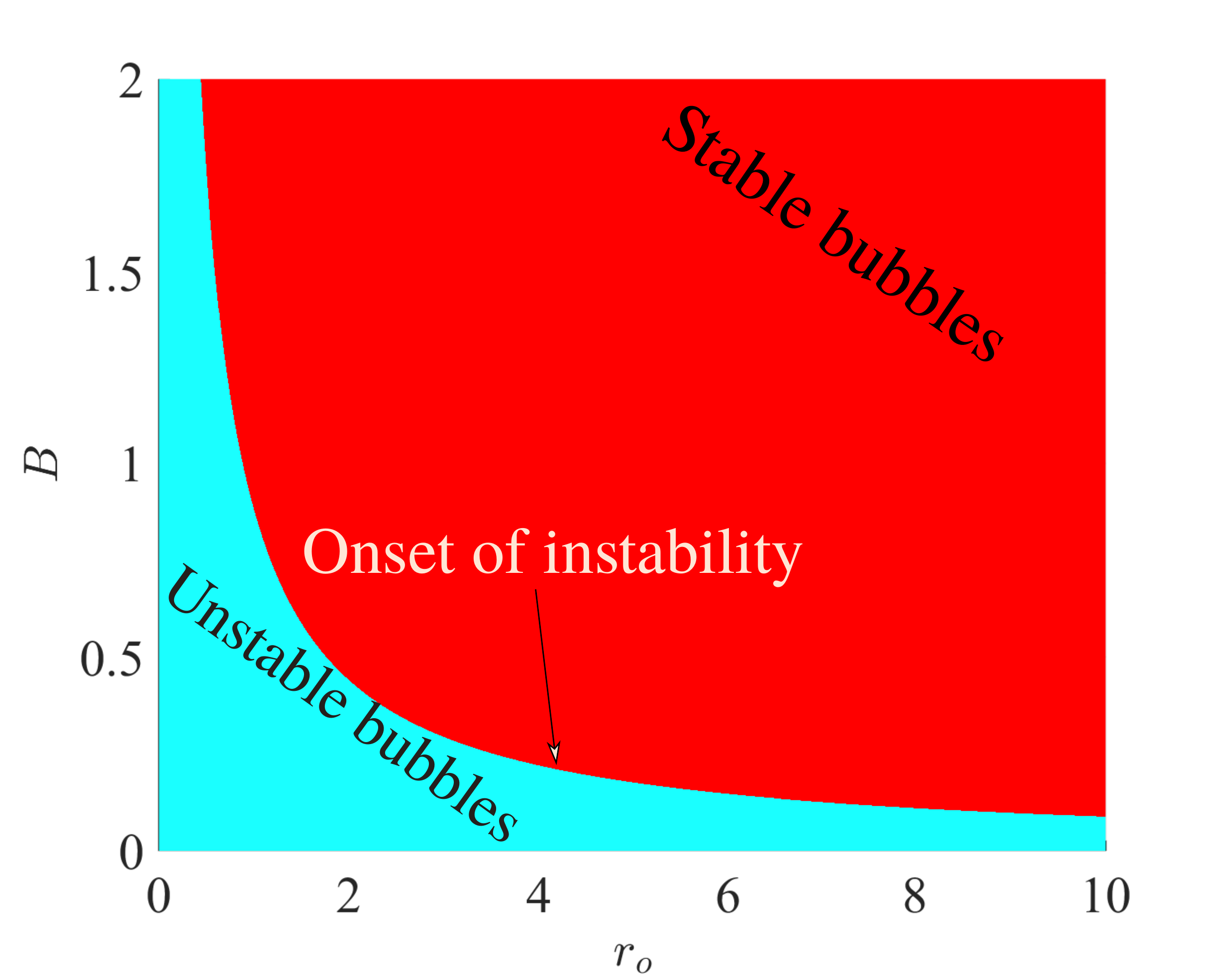}}
  \caption{(Color online) Parameter space $\{r_o,\,B\}$, indicating the regions for stable and unstable bubble fluxons. The boundary between
  both regions is given by the points satisfying the condition $A_c=\sinh(Br_o)$, which gives the onset of instability.}
  \label{fig10}
  \end{center}
 \end{figure}

Figure \ref{fig10} shows a representation of the parameter space that summarises our findings. The blue (red) region denotes combinations of
parameters where bubble fluxons are unstable (stable). Bubble fluxons are stable for $A>1$, and unstable for $A<1$. The onset of instability
that determines the boundary between both regions is given by the condition $\sinh(Br_o)=1=A_c$.  Thus, the stability of bubble fluxons is
described by a single parameter $A$ that couples the two relevant parameters of the system. Given that $r_o$ is related to the
equilibrium radius of the bubble fluxon, and $B$ is related to the injected current, from Fig.~\ref{fig10} is possible to determine the
coaxial dipole current needed for the stabilization of a bubble with a given radius.

The bubble fluxons studied in this work can be regarded as a structural phase transition, where the superconducting phase difference is
driven to a new phase by an instability. In order to produce a transition from one phase to another, a critical germ (a bubble with a
minimum critical radius) is usually considered in the literature for the development of an instability that is energetically above a
nucleation barrier \cite{Gonzalez2006, Gonzalez1999}. In our model, the condition $r_o>1/B$ holds for bubbles with steep walls,
where the bubble decay length is smaller than
its radius. This case corresponds to $A>1$, which is above the onset of instability. Thus, a bubble fluxon can be stabilized by a
coaxial dipole if the steep of the walls is large compared to its radius. Conversely,  the region $A<1$ corresponds to the condition
$r_o<1/B$, which holds when the bubble decay length is larger than its radius. In this case, the bubble is almost collapsed near the origin
and there is no coaxial dipole capable of preventing its annihilation. Previous works have considered the critical germ only in terms of the
minimum radius required in the field configuration $\phi(\mathbf{r}, t)$ \cite{Filippov1992, Gonzalez1999, Khvorostyanov2004, Gonzalez2006,
Scheifele2013}. The results from
our model lead us to the important conclusion that the critical germ is determined by an interplay between its radius and the steepness of
the wall separating the different phases in the system. If the steepness of the wall is increased (decreased), the critical radius decreases (increases).

\section{Conclusions and final remarks}
\label{Sec:Conclusions}

We have investigated theoretically and numerically the stability of bubble-like fluxons in two dimensional Josephson junctions. Although
they are unstable under no external perturbations, we have shown that such structures can be stabilized by the insertion of coaxial
current dipole devices. Using a single parameter $A:=\sinh(Br_o)$ that couples the geometrical properties of the bubble and the coaxial
dipole, we have determined the condition for the formation of stable bubbles. We have predicted theoretically
the creation of oscillating states, the generation of internal mode oscillations and the bubble breakup due to shape mode instabilities.
We have also obtained that the steepness of the walls plays an important role in the critical germ: If the steepness of the wall
is decreased (increased), the critical radius increases (decreases) and a dipole current can stabilize bubbles with large (small) radius.

As a final remark, notice that we have investigated the dynamics of bubbles in the neighborhood of fixed points and separatrices using the
qualitative theory of nonlinear dynamical systems \cite{Guckenheimer1986}. Thus, the external force associated with the dipole current has
not to be exactly Eq.~\eqref{Eq03} to reproduce the same qualitative phenomena. This point is extensively addressed in Ref.~\cite{GarciaNustes2017,
Marin2018, Gonzalez2003, Gonzalez2007}. In our case, any localized force with a fixed point at a certain value of $r$ surrounded by two
extreme values will be qualitatively equivalent to Eq.~\eqref{Eq03}. 

\begin{figure}
 \begin{center}
 \scalebox{0.4}{\includegraphics{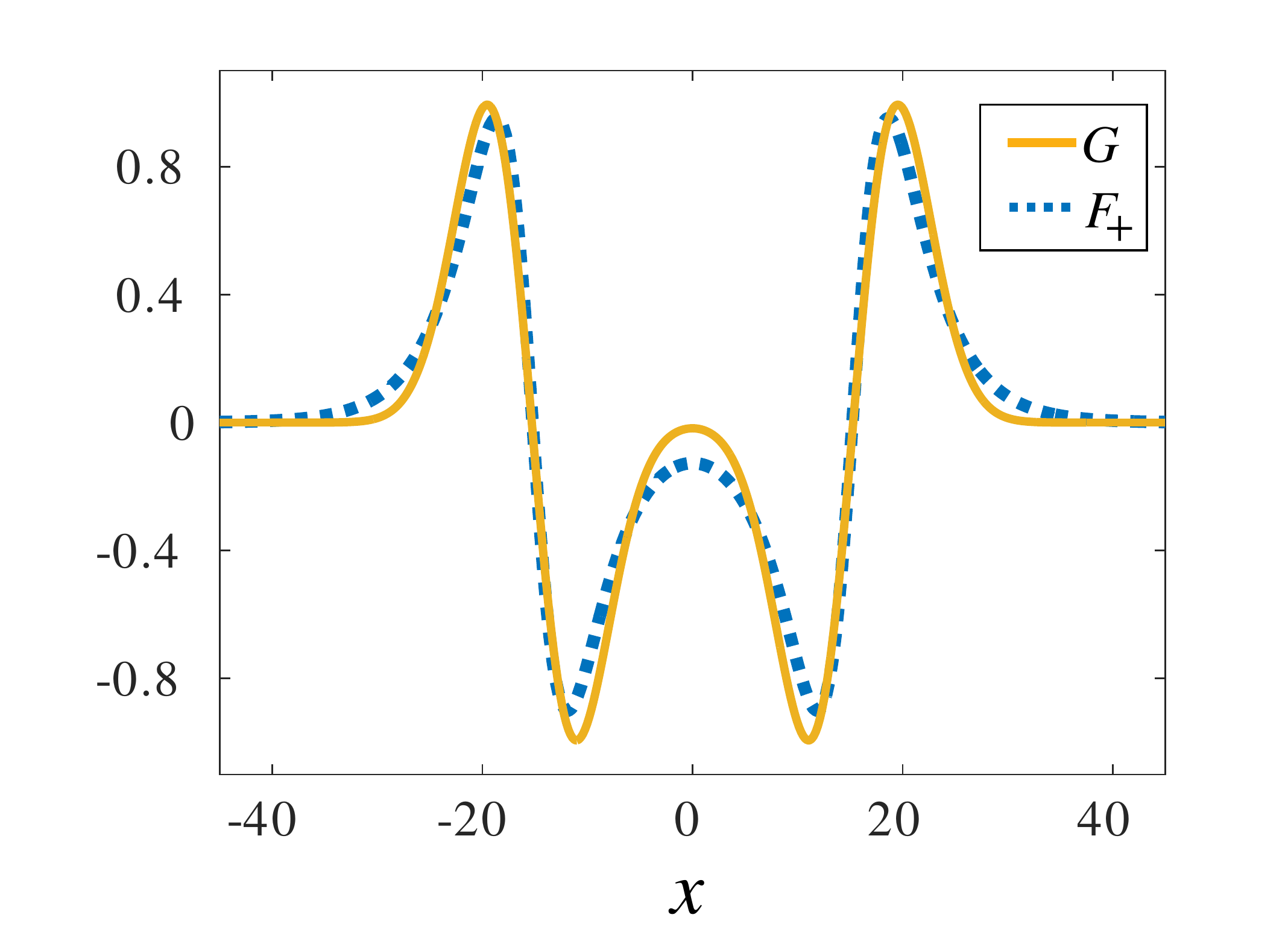}}
 \caption{(Color online) Comparison of topologically equivalent forces for $y=0$. The external force of Eq.~\eqref{Eq03} is depicted in (blue) dashed
 line for $B=0.261$ and $r_f=15.3$. The equivalent force of Eq.~\ref{NewForce} is depicted in (yellow) solid line for $\gamma=3.2$,
 $\sigma=4.2$, $r_o=15.3$, and $a=1.1$.}
 \label{fig12}
 \end{center}
\end{figure}

We conclude that the dynamics of fluxon bubbles in our system is a robust phenomenon. Indeed, the results presented in this article can be
extended to other systems with topologically similar heterogeneities. A particular example is the following superposition of
Gaussian-like functions,
\begin{equation}
 \label{NewForce}
 G(\mathbf{r})=\gamma\left[e^{-\frac{1}{2\sigma^2}(r-r_o+a)^2}-e^{-\frac{1}{2\sigma^2}(r-r_o-a)^2}\right],
\end{equation}
which is depicted in Fig.~\ref{fig12} for the given values of the parameters. It is asserted that $F_{+}$ and $G$ are topologically equivalent
forces for the corresponding values of the parameters. For the force $F_{+}$, we have obtained analytical solutions in full agreement
with the numerical simulations shown in Fig.~\ref{fig08}(a). Notwithstanding, numerical simulations
under the force $G$ show qualitatively the same results.
For the following set of parameters $\left\{\gamma,\,\sigma,\,r_o,\,a \right\}$:
$\left\{0.9,\,0.8,\,30.0,\,0.4\right\}$, $\left\{0.9,\,0.8,\,29.0,\,0.4\right\}$, the force
\eqref{NewForce} reproduces the same results shown in Fig.~\ref{fig02}(b) and 
Fig.~\ref{fig02}(d), respectively. In brief, we have reproduced all the results presented in this
article using several combinations of the parameters of $G$.

In summary, the results of this article provide a robust and general mechanism of control, trapping and breaking of bubble fluxons in JJ
devices. Our analytical results are in good agreement with numerical simulations and suggest a promising way for the storage and transport
of heat and information in quantum information devices.

\begin{acknowledgments}
The authors are grateful to the anonymous referees for their fruitful comments and suggestions that helped to improve
the article. M.A.G-N. thanks for the financial support of Proyecto Interno Regular PUCV 039.306/2018. J.F.M. acknowledges the partial
financial support of CONICYT doctorado nacional No. 21150292 and USA1899-Vridei 041931YZ-PAP Universidad de Santiago de Chile.
\end{acknowledgments}

\bibliography{references}

\end{document}